\documentclass[12pt, a4paper]{article}
 \usepackage{amsmath}
 \usepackage{amsfonts}
\begin{document}
\newcommand{\Dx}{\Delta q}
\newcommand{\ep}{\epsilon}
\newcommand{\qarr}{\stackrel{\cal Q}{\longrightarrow}}
\newcommand{\lc}{L^2 (\Sg_3 (t);\ \bf C;\ b^3 (t){\sqrt \omega (t)} d^3 q)}
\newcommand{\vp}{\varphi}
\newcommand{\1}{{\bf\hat 1}}
\newcommand{\Oc}{O\left(c^{-2(N+1)}\right)}
\newcommand{\im}{{\rm i}} 
\newcommand{\Sg}{\Sigma}
\newcommand{\bgt}{\bigotimes}
\newcommand{\ptl}{\partial}
\newcommand{\Sche}{ Schr\"odinger equation\ }
\newcommand{\Schr}{Schr\"odinger representation\ }
\newcommand{\eu}{$ E_{1,3} $}
\newcommand{\euf}{E_{1,3}}
\newcommand{\rif}{V_{1,3}}
\newcommand{\ri}{$V_{1,3}$ }
\newcommand{\ov}{\overline}
\newcommand{\stc}{\stackrel}
\newcommand{\defst}{\stackrel{def}{=}}
\newcommand{\qstc}{\stackrel{\cal Q}{\longrightarrow}}
\newcommand{\h}{\hbar}
\newcommand{\beq}{\begin{equation}}
\newcommand{\nde}{\end{equation}}
\newcommand{\beqa}{\begin{eqnarray}}
\newcommand{\ndea}{\end{eqnarray}}
\newcommand{\rin}{$V_{1,n}$}
\newcommand{\rinf}{V_{1,n}}
\newcommand{\rn}{$V_n$}
\newcommand{\rnf}{V_n}
\newcommand{\al}{\alpha}
\newcommand{\be}{\beta}
\newcommand{\ga}{\gamma}
\newcommand{\om}{\omega}
\newcommand{\Sch}{Schr\"{o}dinger \ }
\newcommand{\Schs}{Schr\"{o}dinger's \ }
 \newcommand{\lcn}{$L^2 (\rnf;{\mathbb C}; \sqrt\omega d^n q)$}
\newcommand{\drm}{\mathrm d}
\newcommand{\und}{\underline}
\newcommand{\vs}{\vspace*{5mm}}
\begin{titlepage}
\title
{Unfinished History and Paradoxes of Quantum  Potential.\\
 II. Relativistic  Point  of View}
 \author{ E.~A.~Tagirov\\
\small{Joint Institute for Nuclear Research, Dubna 141980, Russia,}\\
 \small {tagirov@theor.jinr.ru}}
\begin{abstract}
This is the second of the two related papers analysing origins and possible explanations of a paradoxical
phenomenon of the quantum potential (QP). It arises in quantum mechanics'(QM) of a particle in the Riemannian
$n$-dimensional configurational space obtained by various procedures of quantization of the non-relativistic
natural Hamilton systems. Now, the two questions are investigated: 1)Does QP appear in the non-relativistic QM
generated by the quantum theory of scalar field (QFT) non-minimally coupled to the space-time metric? 2)To which
extent is it in accord with quantization of the natural systems? To this end, the asymptotic non-relativistic
equation for the particle-interpretable wave functions and  operators of canonical observables are obtained from
the primary QFT objects. It is shown that, in the globally-static space-time, the Hamilton operators coincide at
the origin of the quasi-Euclidean space coordinates in the both altenative approaches for any constant of
non-minimality $\tilde\xi$, but a certain requirement of the Principle of Equivalence to the quantum field
propagator distinguishes the unique value $\tilde\xi = 1/6$. Just the same value had the  constant $\xi$ in
the quantum Hamiltonians arising from the traditional quantizations of the natural systems: the DeWitt canonical,
Pauli-DeWitt quasiclassical, geometrical and Feynman ones, as well as in the revised Schr\"{o}dinger variational
quantization.  Thus, QP generated by mechanics is tightly related to non-minimality of the quantum scalar field.
Meanwhile, an essential discrepancy exists between the non-relativistic QMs derived from the two altenative
approaches:  QFT generate a scalar QP, whereas various quantizations of natural mechanics, lead to PQs depending
on choice of space coordinates as physical observables and non-vanishing even in the flat space if the coordinates
are curvilinear.
   \end{abstract}
\end{titlepage}
\maketitle
\section{Introduction}
In the  accompanying paper under the same  general title and  the  subtitle
 \emph{"I. Non-Relativistic Origin,  History  and Paradoxes."}, to  which I shall refer as ({\textbf I}),
 the main formalisms  of quantization
  of \emph{the  natural Hamilton systems} were analyzed  with interesting  and sometimes paradoxical conclusions.
 The natural systems  are those whose Hamilton functions are    non-uniform quadratic forms in
  momenta $p_a$  with  coefficients $\om^{ab}(q)$ depending  on coordinates
 $ q^{(a)}, \quad a, b,\dots = 1,\dots ,n $ of configurational space $V_n $ :
 \beq
    H^\mathrm{(nat)} (q, p; \om)  = \frac1{2m} \om^{ab}(q) p_a p_b +   V^\mathrm{(ext)}(q).
     \label{Hnat}
  \nde
  \emph{Here and  further, the  notation is used,   which is  standard in   General  Relativity (GR).}
  An  important physical  representative of this class of  systems  is  the particle moving
  in an  external static  gravitational
  field defined  general-relativistically   as the   metric  form of an $n+1$-dimensional
  (Lorentzian) space-time $V_{1,n}$ in the normal  Gaussian system of coordinates $\{x^0\equiv ct, \, q^{(a)}\}$:
 \beq
    ds_{(g)}^2 = g_{\al\be}(x) dx^\al dx^\be = c^2 dt^2 - \om_{ab}(q) dq^a dq^b , \quad
    \al,\be, \dots = 0, 1,\dots n ; \label{ds}
    \nde
    Then, this  construction   is  \emph{a foliation of   $V_{1,n}$ ( frame of reference)}
     by the  normal geodesic translations  of  any    space-like hypersurface
   \beq
   \Sg =\{x \in V_{1,n},\ \Sg(x)= const, \ \ptl^\al\Sg(x) \ptl_\al \Sg(x)>0 \} \label{sg}
   \nde
   the  interior  geometry  of  which is  that  of $V_n$.  If a metric
  tensor $\om_{ab}$ does  not  depend on  $t$,  \ $V_{1,n}$ is  a  globally static space-time.

  Analysis of various  quantization procedures  of the generic  natural system  in \cite{TagI}
  has shown that   the  resulting  non-relativistic QMs  of a particle  do not reconcile with the basic principles
  of GR , namely, the Principles  of General Covariance and  of Equivalence, owing to inevitable  appearance
  of  QPs in  the Hamilton   operators or propagators. In the formers,  these QPs are  not invariant
  (not  scalars) with respect to general
  transformations of coordinates    ${q^\prime}^a = {q^\prime}^a(q)$ and they single out persistently
   the  potential term:
     \beq
          V^\mathrm{(qm)}(y) = - \frac{\h^2}{2m} \cdot \frac16 R_{\mathrm(\om)}(0) + O(y), \label{Vqmy}
               \nde
    at the  origin of the quasi-Euclidean  (normal Riemannian)
  coordinates $y^a$, where   $R_{\mathrm(\om)}(q)$ is  \emph{the  scalar curvature} of  $V_n$.
  It   contradicts formally to  the  Principle of  Equivalence (PE) in  S.~Weinberg's formulation
  \cite{Wein} quoted also in \cite{TagI}, Section 3.

  In view of  this  paradoxes, we shall consider  now an alternative  approach to construction the non-relativistic QM
  \emph{in the   globally static}  $V_{1,n}$,
  which  starts from  the  general-relativistic  quantum  theory of  a neutral scalar field    and
       produces  a non-relativistic QM as the  limit for $c^{-1}\rightarrow 0 $ of the
  one-quasi-particle sector of   an    appropriate Fock  representation. The  initial theory is  general-covariant
   and  extraction  of  QM  from  it  is  covariant with  respect to  transformations of the  spatial  Gaussian
     coordinates $q^a$. As concerns PE in quantum theory,  the  field-theoretical approach shows, in which sense it
      is  satisfied   on the  relativistic level,  and    originates the term  (\ref{Vqmy}) in the  non-relativistic QM.

     The paper is  organized as  follows. In  Section 2 , a brief exposition of the classical theory of
      scalar field in  $V_{1,n}$ non-minimally coupled to  the  metric  is  given.   In Section 3
       and relation of  the  energy-momentum  tensor in  the
     conformal covariant version of the theory to  the Dirac  scalar-tensor  theory  of  gravitation is  shown.
     In  Sections 4 - 6, the  Fock representations of  quantum theory  of  the  field  is  constructed and
     and  relation to PE of  the  structure on the light conoid  of  the propagator  is  considered.
      Restriction to  the  time-independent    (globally static) case, which is necessary for
      comparison with conclusions of \cite{TagI},
      is considered in  Sections 7-8. A logical chain of  conclusions of the both  papers is given in Section 9.

\section{Scalar Field  in  Riemannian space-time,  conformal \\ covariance and  Principle of  Equivalence}
  Thus, we start with the (classical)   real  scalar   field  $\varphi(x), \quad  x \in V_{1,n}$,
  which  satisfies  to  the so called  non-minimal  generalization   of   the   standard  Klein--Gordon--Fock
  equation:
  \beq
  \Box \varphi + \tilde\xi\, R_{(g)}(x)\,  + \left(\frac{mc}{\h}\right)^2 \varphi = 0, \qquad
  \Box \defst g^{\al\be}\nabla_\al\nabla_\be \equiv (- g)^{-\frac12}\ptl_\al \left((- g)^\frac12 g^{\al\be}\ptl_\be
  \right).
   \label{eq}
    \nde
    \emph{Notation here  and  in sequel is}
    \begin{itemize}
     \item $\nabla_\al $ is  the covariant derivative in $V_{1,n}$;
     \item
      $R_{(g)}  = g^{\al\be}R^\gamma_{(g)\al\gamma\be}$  is the  scalar curvature of $V_{1,n}$ and  the
      Riemann-Christoffel cuvature tensor is  determined so  that
     $(\nabla_\al\nabla_\be - \nabla_\be\nabla_\al)f_\gamma   = R^\delta_{(g)\gamma\al\be}f_\delta $ for any twice
     differentiable 1-form $f_\gamma (x)$;
     \item
     $\tilde\xi \equiv const$   is  \emph{a (dimensionless) parameter  of  non-minimality}
   of  the  coupling of   $\varphi (x)$  to the external  gravitation represented  by  the  metric  tensor
    $ g_{\al\be}(x)$; the value $\tilde\xi = 0 $  corresponds to  the  minimal coupling    traditionally adopted
    in theoretical physics    up  to  the  end  of 1960s.
   \end{itemize}

     Among the arbitrary values of  $\tilde\xi$,  there is a distinguished   value
      \beq
    \tilde\xi = \tilde\xi_\mathrm{(conf)}(n) \defst\, \frac{n-1}{4n} \label{conf}
   \nde
  for   which  eq.(\ref{eq}) is  \emph{asymptotically conformal covariant} for  $m \rightarrow 0 $, that is, if
     $\varphi(x), \, x\in V_{1,n}$ is   a solution of eq.(\ref{eq}) with  $m=0$,  then
     $\tilde \varphi (x) \defst \Omega^{\frac{1-n}2} (x)\, \varphi (x),\  x \in \tilde V_{1,n}$ , is a solution
     of the same       equation
     in  $\tilde V_{1,n}$ whose metric tensor is $\tilde g_{\al\be} (x) = \Omega^2(x)\, g_{\al\be} (x)$ and
   $\Omega(x)$ is
 an arbitrary sufficiently smooth  function.\footnote{This property of eq.(\ref{eq}) was first pointed out
by   R.~Penrose \cite{Pen} but  only  for  $n=3$ and  with  no  consideration  of  physical consequences. Detailed
study    of   properties   and quantization of  $\varphi(x)$ satisfying eq.(\ref{eq}) with
   $\tilde\xi = \tilde\xi_\mathrm{(conf)}(n) $   for arbitrary dimension $n$ was done in \cite{ChT}(see also \cite{BT}). Soon after
   that,   eq.(\ref{eq}) attracted a serious attention in theoretical physics  and cosmology in
    its  asymptotically conformal covariant as well  general  non-minimal  forms; see the analytic review \cite{Far}
     by V.~Faraoni on the
   role  of  different values of $\tilde\xi$  in  the  generalized  inflation  models in cosmology.}
   Conformal  covariance  ensures \emph{conformal  invariance} of eq.(\ref{eq}) and  corresponding
   conservation  laws     if $ V_{1,n}$ under  consideration    admits  a group  of  conformal  isometries (motions).

   The term $\tilde\xi\, R_{(g)}(x)$  in eq.(\ref{eq}) again,  as in the \Sch equation with QP,   causes  the
   question  on PE (see  the  formulation by S.~Weinberg \cite{Wein}
   reproduced also in  \cite{TagI})      since the term does not
    disappear in the  quasi-Cartesian coordinates with  the  origin at
    $x$   if $R_{(g)}(x)\neq 0$.  Some answer  on  the  question  gives an  investigation
   of structure of singularities  of  the   Green functions  for the field  equation (\ref{eq}). First, in 1974,
 S.~Il'in and  the present  author \cite{IT} had shown that for
\beq
 \lim_{x\rightarrow  x^\prime}\left\{\bar{\Bbb G}_{V_{1,3}}(x,\, x^\prime ; \tilde\xi) -
 \bar{\Bbb G}_{E_{1,3}}(\Gamma(x,\, x^\prime))\right\} = \frac{\theta(\Gamma(x,\, x^\prime))}{8\pi}(\tilde\xi-
\frac{1}{6})R_{(g)}(x^\prime)\
 \nde
 where $ \bar{\Bbb G}_{V_{1,3}}(x,\, x^\prime ; \tilde\xi)$ is  the classical    Green function  in $V_{1,3}$ and
 $ \Gamma(x,\, x^\prime)$  is  the geodesic  interval    between  $x,\, x^\prime $.
 Thus,  singularities of $\bar{\Bbb G}_{V_{1,3}} (x,\, x^\prime ; \tilde\xi)$ on  the  light conoid
         $\Gamma(x,\, x^\prime) = 0 $ (the
    locus of  isotropic geodesics, emanated from   $x^\prime$) are the same as in the  tangent  space $E_{1,3}$,
    "a locally inertial coordinate system" in  Weinberg's formulation of PE, see  \cite{TagI}.
    Thus,  PE is satisfied in this   sense in the  classical field theory with $\tilde\xi =\frac16 $  and $n=3$ (The direct
    recalculation in $V_{1,n}$ shows that the same property takes  place also for  arbitrary $n$).
     Unfortunately,  the  authors  of \cite{IT} had  not recognized sufficiently   the  significance of their  result
     for justification of PE for  eq.(\ref{eq}). Therefore,  it  is  not suprising that  much  later,
       Sonego and Faraoni \cite{SF} have   reproduced, in  fact,  the same result but as a  verification  of PE.

 Generalization  of  this verification to the quantum theory given by A.~A.~Grib and E.~A.~Poberii \cite{gp} will
 be noted in Section 6 after quantization of  field  $\varphi$.

\section{Energy-momentum  tensor and   Dirac scalar-tensor\\ theory}

     Eq.(\ref{eq}) is the  unique linear   covariant scalar field equation if
   one  introduces  no new  dimensional constant into  the  theory \cite{Tag4}.
   It follows from variation of   $\varphi $ in  the  functional of action
    \beq
     {\cal A}\{g_{..}(.), \varphi(.); \tilde\xi\} \defst \int L (x) (-g)^{\frac12}\drm^{n+1} x ;\qquad
   L  \defst \frac12  \ptl^\al\varphi \ptl_\al \varphi - \frac12 \left(\left(\frac{mc}\h\right)^2
   + \tilde\xi R_{(g)}\right)\varphi^2.\label{A}
     \nde
     Its variation by $g_{\al\be}(x)$ gives  the (metric) energy-momentum tensor
      \beqa
T_{\alpha\beta} (x; \tilde\xi)&\defst& \frac{\delta{\cal A}\{g_{..}(.), \varphi(.); \tilde\xi\}}{\delta g^{\al\be}(x)} \nonumber \\
 &=&\vp_\alpha \vp_\beta - L g_{\al\be} -  \tilde\xi ({R_{(g)}}_{\alpha\beta } - \frac12 R_{(g)} g_{\alpha\beta} +  \nabla_\alpha
\partial_\beta - g_{\alpha\beta }\Box ) \varphi^2,  \label{tem}
 \ndea
For  solutions  of eq.(\ref{eq}), one has
 \beq
  T (x; \tilde\xi)\defst g^{\alpha\beta} T_{\alpha\beta} (x; \tilde\xi)
   = \left(\frac{mc}\h\right)^2 \varphi^2   + n \left(\tilde\xi -\tilde\xi_\mathrm{(conf)}(n)\right)
   \left(\varphi^\al\varphi_\al- 2\left(\tilde\xi R_{(g)}+ \left(\frac{mc}\h\right)^2 \right)\varphi^2 \right),
         \label{Tr}
 \nde
 and consequently
 \beq
  T \left(x; \tilde\xi_\mathrm{(conf)}(n)\right ) = \bigl(\frac{mc}{\h}\bigr)^2 \varphi^2, \label{TT}
 \nde
i.e.,  it has the  property which  is  inherent  also  for fields with  spin 1/2 and 1 and  which provides all
these fields with the asymptotic  conservation laws corresponding to conformal isometries (if  any) when $ m
\rightarrow 0$. Note also, that    $T_{\alpha\beta} (x; \tilde\xi)  \neq  T_{\alpha\beta}(x; 0)$ even in $E_{1,n}$
if $\tilde\xi \neq 0 $.

  Tensor $T _{\alpha\beta} (x; \tilde\xi_\mathrm{(conf)}(3))$ has been re-discovered  later and called
  \emph{"a  new energy-momentum
  tensor"} by Callan, Coleman and Jackiv \cite{CCJ}.  They had  postulated \\
   $T_{\alpha\beta} (x; \,\tilde\xi_\mathrm{(conf)}(3)) $   in the  form
   of   eq.(\ref{tem}) for the particular  case of  $E_{1,3}$ and   generalized  it afterwards
     for  $V_{1,3}$. Their   reasoning  is evidently   an inversion  of  the
     straightforward  general-relativistic   approach with the requirement of  the  conformal symmetry  in \cite{ChT}.

     More interesting  is  that,  in 1973, Dirac\cite{Dir}  formulated  a  scalar-tensor theory of  gravitation in  relation  with
     his famous  hypothesis on  large  numbers.  For $n=3$ and  $\tilde\xi = 1/6$,  the integral  $\cal A $   is  just the
     gravitational (geometrical) part of
     the action integral of   the Dirac theory \cite{Dir}, formula (5.2) there.    (The full Dirac action integral includes
       also  the  electromagnetic $F_{\mu\nu}F^{\mu\nu}$ and  non-linear $const \cdot\vp^4 $ terms.) Therefore, our
       $T_{\alpha\beta} (x; 1/6)$ is just the  left-hand side of the   scalar-tensor  Dirac equation .
       In fact, Dirac had been motivated by   simplicity of the  trace $T (x; \tilde\xi)$, eq.{\ref{TT}}, when
       $\tilde\xi =\tilde\xi_{\mathrm(conf)}(3)\equiv\frac16 $.
         However, we see that the  same reasoning  is  correct  for any $n$ and, thus, the  Dirac  theory can
         be generalized to  any  $V_{1,n}$ as a conformal-covariant one.  In fact,   the theory based on
        the  action  integral  ${\cal A}\{g_{..}(.), \varphi(.);  \tilde\xi_\mathrm{(conf)}(3)\}$ is used for construction
        of so called  conformal cosmology, an altenative to the  standard model,  and  applied  to  fit recent  data
        on distant supernovae taken  as standard candles, \cite{Per} and  references therein. Thus, determination
        of value of $\tilde\xi$ acquires a "practical"  interest.

   \section{Quantization of the scalar field  in the general Riemannian space-time}

  Now, the  quantum  theory  of  the  field $\varphi(x), \  x \in V_{1,n}$
   (denoted as QFT in sequel) will be formulated to  extract from
  it a structure similar to the non-relativistic QM considered  in \cite{TagI}.  The  program of construction of a
  particle-interpreted   Fock representation for quantum  field $\check\varphi(x), \  x \in V_{1,n}$,
    has been fulfilled
  in \cite{Tag2} with use of   formulations from \cite{Grib}, Chapter 2, and \cite{dav}, Chapter 3,
  (\emph{"check" over    symbols will denote operators
  in the  Fock spaces $\cal F $}). Here, the main points  of that  program with some  improvements including
  a  consideration of PE in QFT  will be reproduced in the following  four   sesections
    for a consecutive  statement of the  problem and conclusions.

  The program    starts  with   complexification
  $\Phi_c\,= \,\Phi \otimes {\Bbb C}$,  of the space  $\Phi $ of solutions to eq.(\ref{eq})
  and  a subspace $\Phi_c^\prime \subset \Phi_c $    such that
 \begin{equation}
    \Phi_c^\prime\,=\,  \Phi^- \oplus  \Phi^+  \label{c}
 \end{equation}
 where $  \Phi^\pm  $ are supposed to be mutually complex conjugate spaces. They are selected so that   the
 conserved  (i.e. independent
on choice  of $\Sg$)  Hermitean sesquilinear form
  \begin{equation}
 \{ \varphi_1, \,\varphi_2 \}_\Sigma \defst  i\int_\Sigma \drm\sigma^\alpha
\left(\ov {\varphi_1}(x)\,\ptl_\alpha \varphi_2 (x)\ - \, \ptl_\alpha \ov {\varphi_1}(x)\,{\varphi_2}(x)\right),
\label{spr}
\end{equation}
be positive (negative) definite  in $\Phi^+ (\Phi^-)$, where   $d\sigma^\al$ is  the normal  volume element of a
Cauchy hypersurface $\Sg$ induced by the metric of \rin and determined  for an arbitrary vector field $f^\al (x)$
and arbitrary interior coordinates ${q^a}$ on $\Sg$ by  relation
 \[f_\al \drm \sigma^\al = (- g)^{\frac12}
 \left|\begin{array}{cccc}
    f^0 & f^1 & \dots & f^n \\
   \frac{\ptl x^0}{\ptl q^1} \drm x^0 & \frac{\ptl x^1}{\ptl q^1}\drm x^1 & \dots & \frac{\ptl x^n}{\ptl q^1} \drm x^n \\
      . & . & \dots & .\\
   \frac{\ptl x^0}{\ptl q^n} \drm x^0 & \frac{\ptl x^2}{\ptl q^n}\drm x^1 & \dots & \frac{\ptl x^n}{\ptl q^n}
   \drm x^n
 \end{array}\right| \]
   The form (\ref{spr}) can thus be considered as an inner product in  $ \Phi^- $ providing the last
   with a pre-Hilbert structure.

It is  clear that  bi-partition (\ref{c}) of $\Phi_c$  can be  done by  an infinite set of  ways. In $E_{1,n}$ and
the globally static $V_{1,n}$, there is  a discriminated   bi-partition  by the  positive- and negative-frequency
solution  owing to  existence of the conserved positive definite observable of energy. However, for a time being,
the generically time-dependent  $V_{1,n}$ makes sense to  be considered.

Let, further, $\{\varphi (x;\, A\}\subset \Phi^+ $ be a basis enumerated by  a multi--index $A$ , which has values
on a set $\{A\}$ with a measure $\mu (A)$, and  orthonormalized with respect to the inner product (\ref{spr}).
 Then,
\begin{equation}
 \check \varphi (x) = \int_{\{A\}}  \drm  \mu(A) \left(\check c^+(A)\,
\ov\varphi (x;\,A)\, +\, \check c^-(A)\, \varphi (x;\,A)\right)\, \equiv \,\check\varphi^+ (x) \,+
\,\check\varphi^- (x),
  \label{a+}
\end{equation}
with  the operators $\check c^+(A)$ and $\check c^-(A) $ of creation and annihilation of the  field modes
$\varphi^- (x;\, A) \in \Phi^- $ (\emph{or, of the quasi-particles}), which  satisfy the canonical commutation
relations
 $$
 [\check c^+(A),\,\check c^+(A')] = [\check c^-(A),\,\check c^-(A')] = 0,\ \int_{\{A\}}  \drm \mu(A)\, f(A)\,
 [\check c^-(A),\, \check c^+(A')] = f(A')
 $$
  for any appropriate  function $f(A)$. They act  in the Fock space $\cal F$ with the cyclic vector $ |0> $ (\emph{the
quasi-vacuum}) defined by  equations
\begin{equation}
\check c^-(A)\, |0> = 0.  \label{vac1}
\end{equation}
 The  conservation  property of  the  "scalar product" (\ref{spr}) allows to  consider the  basis as defined
 on the space of the  Cauchy  data  on  a  concrete
hypersurface $\Sg  $,   but \emph{the different  choices  of  $\Sg $ determine
 different Fock spaces  $\cal F$ which  are,  in general, unitarily  uneqvivalent},  see, e.g.,
\cite{Grib}. Correspondingly, $|0> \equiv |0;\Sg>$  and ${\cal F}\equiv {\cal F} \{\Sigma\} $. Then, operators of
the basic  observables  in ${\cal F} \{\Sigma\}$ can be defined as follows.

\emph{The operator of number of  quasi-particles}

\beq \check{\cal N}\{\check \varphi;\,\Sg\} \defst
 i\int_\Sg  \drm\sigma^\alpha\, (\check\varphi^+\, \ptl_\alpha \check\varphi^-
- \ptl_\alpha \check\varphi^+\, \check\varphi^-)   \defst  \int_\Sg  \drm\sigma (x) \check N (x), \quad
  \drm\sigma  \defst \frac{\ptl_\al \Sigma \drm\sigma^\al}{(\ptl^\al \Sigma \ptl_\al \Sigma)^\frac12}.  \label{n}
\nde
 \emph{The operator of projection of momentum of field $\check \varphi (x)$ on a given vector field}
 $K^\alpha(x)$:
\begin{equation}
\check {\cal P}_K \{\check\varphi ; \, \Sg \} \, \defst\,  :\int_\Sg \drm\sigma^\alpha \ K^\beta T_{\alpha\beta}
(\check\varphi):, \label{pk}
\end{equation}
where and in sequel \emph{the colons denote  the normal product of operators} $c^\pm_\Sg $ .

To define \emph{a QFT- prototype}
 $\check{\cal Q}^{(a)} \{\check\varphi;\, \Sigma\} , \quad a, b, \dots  = 1, \dots n$ of  non-relativistic QM
 position operators $\hat q^a$ which played a
 basic  role  in \cite{TagI},  introduce first $n$ position-type
 functions   $q^{(a)}(x), \  x \in  V_{1,n}$  which are defined  in \cite{Tag2},  Section 2, in terms of
   fibre bundles. Consideration in the  present paper is  restricted  by   the traditional
   conjecture in  theoretical physics
 that  $ V_{1,n}$ is  a  trivial manifold. (It is  equivalent in physics to assumption that  only
 local  manifestations  of
 the curvature are taken  into  account.)      Then, it is   sufficient to  introduce     $q_\Sg^{(i)}(x)$ are
 \emph{scalar functions} of $x^\al $\emph{ w.r.t.} general  transformations  $\tilde x^\al =  \tilde x^\al (x)$,
 which  satisfy the  conditions
\begin{equation}
 \bigl. \ptl^\alpha\Sg \  \ptl_\alpha q_\Sg^{(i)}\,\bigr|_\Sg =  0 , \qquad
 \bigl. \mbox{rank}\|\ptl_\alpha q_\Sg^{(i)}\|\, \bigr|_\Sg = 3,    \label{q}
\end{equation}
So, they  define a point on the Cauchy hypesurface $\Sg = \{ x\in\rif\,|\, \Sg(x) = const\}$. Their restrictions
on $\Sg$ can serve as internal coordinates on it.

 Assuming that the corresponding QFT--operators $\check{\cal Q}^{(i)}\{\check\varphi;\, \Sg\}$ have the
same structure  as the operators $\check{\cal N}$ and $\check{\cal P_K}$  introduced above, let us  impose the
following conditions on them             :
\begin{enumerate}
\item
$\check{\cal Q}^{(i)} \{\check\varphi;\, \Sg\}$ should be local sesquilinear Hermitean forms in the operators
$\check\varphi^\pm (x)$, and linear functionals of  $q_\Sg^{(a)} (x) $ expressed as invariant integrals over
$\Sg$.
\item
$\check{\cal Q}^{(i)}\{\check\varphi;\, \Sg\}$ should not contain derivatives of $q_\Sg^{(i)} (x) $.
\item
$\check{\cal Q}^{(i)}\{\check\varphi;\, \Sg\}$ should lead to the operator  of multiplication by $q_\Sg^{(i)} (x)$
in the configuration  space of the standard non-relativistic  QM, i.e. for $c^{-1} = 0 $ .
\end{enumerate}
These conditions lead apparently to the following unique set of $n$ operators  $\check{\cal Q}^{(i)}$ on  $\cal
F$:
\begin{eqnarray}
\check{\cal Q}^{(i)}\{\check\varphi;\, \Sg\} &\defst & i
 \int_\Sg \drm\sigma^\alpha (x)\ q_\Sg^{(i)}(x)\, \left(\check\varphi^+(x)\
\ptl_\alpha\check\varphi^- (x)\,
- \, \ptl_\alpha\check\varphi^+ (x)\,\check\varphi^- (x)\right) \nonumber \\
&\equiv & \int_\Sg \drm\sigma (x)\, q_\Sg^{(i)}(x)\, \check N (x).\label{Q2}
\end{eqnarray}
This  definition, in a  certain sense,   leads to a generalization for \ri of the known Newton--Wigner operator of
the  Cartesian  coordinate  operators  as it is shown in  \cite{Tag2}, Section 6.

 \section{One-quasi-particle subspace of  Fock  space}

 A normalized one-quasi-particle state vector in ${\cal F} \{\Sg\}$ is
\begin{equation}
  |\varphi> \defst \{\varphi,\,  \varphi \}_\Sg^{-1/2}
\int_{\{A\}}\drm\mu (A)\,\{\varphi (.\,; A),\,\varphi (.)\}_\Sg\ \check c^+(A)\,|0; \Sg>.
  \label{phi}
\end{equation}
It determines  the field configuration
\[
 \Phi^- \,\ni\,\varphi (x)
= \int_{\{A\}} \drm\mu (A)\,\{\varphi (.\,;\, A),\,\varphi (.)\}_\Sg\ \varphi (x;\,A).
\]
Obviously $<\varphi | \varphi> = 1$.

  Consider matrix elements of  operators
 $\check {\cal N} (\check\varphi;\,\Sg),\ \check {\cal P}_K (\check\varphi; \,\Sg)$ and $\check{\cal Q}^a\{\check\varphi;\,
\Sg\}$ between  two such states $|\varphi_1>$  and $|\varphi_2>$. Simple calculations  with  use of Eqs.(\ref{n}),
(\ref{tem}), (\ref{Q2}) and (\ref{phi}) give:
\begin{equation}
  <\varphi_1|\,  \check {\cal N} (\check \varphi;\,\Sg)\, |\varphi_2>
= \frac{\{\varphi_1,\, \varphi_2 \}_\Sg}{\{\varphi_1,\, \varphi_1\}_\Sg^{1/2} \{\varphi_2,\,
\varphi_2\}_\Sg^{1/2}} , \label{N1}
\end{equation}
\begin{equation}
<\varphi_1| \check {\cal P}_K (\check\varphi; \Sg)|\varphi_2> = \frac{P_K (\varphi_1,\, \varphi_2;\, \Sg)}
{\{\varphi_1,\, \varphi_1 \}_\Sg^{1/2} \{\varphi_2,\, \varphi_2\}_\Sg^{1/2}} \label{pk1}
\end{equation}
where
\begin{eqnarray}
P_K (\varphi_1,\, \varphi_2;\, \Sg) = \h\int_\Sg \, \drm\sigma^\alpha \,
 \biggl(\ptl_\alpha \ov\varphi_1\, K^\beta\ptl_\beta \varphi_2
+ K^\beta \ptl_\beta\ov\varphi_1\, \ptl_\alpha \varphi_2
\qquad\qquad\qquad\qquad\qquad\qquad & & \nonumber \\
\qquad  -\,  K_\alpha  \left(\ptl_\beta \ov\varphi_1\, \ptl^\beta \varphi_2 -
\left(\left(\frac{mc}{\hbar}\right)^2 + \tilde\xi R_{(g)}\right)\ov\varphi_1\,\varphi_2\right) \nonumber \\
- \biggl. \tilde\xi\, \int_\Sg \, \drm\sigma^\alpha \,(\tilde K_{\alpha\beta} \ptl^\beta - \nabla^\beta \tilde
K_{\alpha\beta})\, (\ov\varphi_1 \varphi_2)\biggr) \label{pk2}
\end{eqnarray}
 where
\begin{equation}
\tilde K_{\alpha\beta} \defst \nabla_\alpha K_\beta + \nabla_\beta K_\alpha - \nabla K \, g_{\alpha\beta}
 \label{tild},
\end{equation}
and
\begin{equation}
<\varphi_1|\,\check{\cal Q}^a\{\check\varphi;\, \Sg\}\,|\varphi_2> = \frac{\{\varphi_1,\, q_\Sg^{(a)} \varphi_2
\}_\Sg}
  {\{\varphi_1,\, \varphi_1 \}_\Sg^{1/2}
\{\varphi_2,\, \varphi_2\}_\Sg^{1/2}} \label{Q1}
\end{equation}
These matrix elements are    sesquilinear functionals    of two functions $\varphi_1 (x),\, \varphi_2 (x)\in
\Phi^- $  which are obviously Hermitean  in the sense that,  given  a functional
 ${\cal Z}(\varphi_1,\,\varphi_2;\, \Sg)$, the following equality takes place:
\begin{equation}
{\cal Z}(\varphi_1,\,\varphi_2;\, \Sg) = \ov{{\cal Z}(\varphi_2,\,\varphi_1;\, \Sg)}.  \label{z}
\end{equation}
\section{Principle  of Equivalence in quantum  field theory}
  Representation (\ref{a+})  of  the quantum  field $\check \phi$ allows to  obtain  the  causal Green function
  (or, the propagator of the quasi-particle).
   \beqa
  {\Bbb G}^\mathrm{(causal)}_{V_{1,n}} (x,\, x^\prime ; \tilde\xi)
 &\defst& \frac1i  <0; \Sg| T(\check\varphi(x)\check\varphi(x^\prime))|0; \Sg> \label{T}\\
   &=&  \bar{\Bbb G}_{V_{1,n}} (x,\, x^\prime ; \tilde\xi) +
   \frac{i}2 {\Bbb G}^\mathrm{(1)}_{V_{1,n}} (x,\, x^\prime ; \tilde\xi),   \label{caus}
     \ndea
  where $T$ denotes  the  chronological product and ${\Bbb G}^\mathrm{(1)}$  is the Hadamard elementary solution
  for the  field equation (\ref{eq})
  which is determined up  to  a  regular solution of (\ref{eq})  $w(x, x^\prime)$ satisfying the
  initial condition $ w(x, x^\prime) \rightarrow 0\ \mbox{for}\ x \rightarrow x^\prime $.
   Since, in  general,   the  definition of quasi-particles  and quasi-vacuum  depend  on choice of the  initial
   Cauchy hypersurface $\Sg_0$,  the bi-scalar  $w(x, x^\prime)$ does, too,  according  to definition (\ref{T}),
    and  determines
   creation and  annihilation  of  the newly determined   quasi-particles  when   $\Sg_0$ (system  of  reference)
   is changed.

   Contrary to \cite{IT},  A.~A.~Grib and E.~A.~Poberii \cite{gp}   studied  both terms
  in  eq.(\ref{caus}) together and  have  obtained  that
 \beqa
 &&\lim_{x\rightarrow  x^\prime}\left\{{\Bbb G}^\mathrm{(1)}_{V_{1,3}}(x,\, x^\prime ; \tilde\xi)\right. -
 \left.
{\Bbb G}^\mathrm{(1)}_{E_{1,3}}(\Gamma(x,\, x^\prime)) \right\}\nonumber\\
&& \qquad \qquad = \lim_{x\rightarrow  x^\prime}\left\{\frac1{8\pi^2 }  (2\gamma + \ln |m^2 \Gamma (x,\,
x^\prime)|) (\tilde\xi- \frac{1}{6})R_{(g)}(x^\prime) + w(x,\, x^\prime)\right\}. \nonumber \ndea
  Thus, they   have shown  directly that  the  quantum  Green function    supports  PE if $\xi=1/6$.  All
  the works  mentioned above
  are restricted by  the case of $n=3$ but re-calculation for abitrary $n$ leads to the  same result amd
  therefore we come to an important conclusion that \emph{the (asymptotic) conformal covariance  and PE are in accord
  only for  n=3 and  thus the dimensionality of our real space is  distinguished by that}.

\section{From quasi-particles to  a quantum point-like  particle}
Now, our  main  aim  is   to extract a counterpart to  non-relativistic QM of  the natural mechanical systems,
that had been  considered in \cite{TagI}, from the ambiguous relativistic  one-quasi-particle structure just
described, and to compare these two QMs. The space $\Phi^-$ so discriminated  could be interpreted on a sufficient
physical basis as the space of wave functions of particles instead of the ambiguous notion of a quasi-particle. In
\eu \ and globally static space-times, there  exists an unique decomposition (\ref{c}) such that  an irreducible
representation of the space-time symmetry is realized on $\Phi^-_\Sg $ but, even in these exceptional cases, one
should   restore  the quantum-mechanical operators on
 $ L^2 (V_n; {\mathbb C}; \drm\sigma)$ of canonical observables of coordinates $q^a$ and of  momenta
$p^a$ cojugate to  them; this  is not a completely  evident task. In sequel
 \emph{the operators in  $L^2 (V_n; {\mathbb C};  \drm\sigma)$  and  its analogs are denoted by  "hat" on top };
 and the  superscript "(ft)" denotes  objects of the field-theoretical origin. All "hatted" operators act
 along the hypersurface  $\Sg \ni x $ or  its normal geodesic translations $S_\Sg =const$  are expressed  in  terms of
    projections of  covariant derivatives $\nabla_\al$ onto  these hypersurfaces:
\begin{equation}
 D_\alpha \defst h_\alpha^\beta \nabla_\beta, \quad
h_\alpha^\beta  \defst  \delta_\alpha^\beta - \ptl_\al S_\Sg \ptl^\beta S_\Sg , \label{D}
\end{equation}
(i.e. $ h_{\alpha\beta}$  is \textit{ the tensor of projection on $S_\Sg$}). I recall that,  up for  a time being,
we consider non-static $V_{1,n}$  for generality.

 Our first task  is to   construct  a map
 \beq
 \Phi^-_\Sg \ni \varphi \longrightarrow  \psi(x)\in L^2 (S_\Sg; {\mathbb C}; \om^{1/2}\drm^n q)\label{map}
 \nde
so that eq.(\ref{eq}) would generate     \Sch--DeWitt-type   equation, eq.(17) in \cite{TagI}   in terms
   of $\psi(x)\in L^2(\Sg; {\mathbb C}; \om^{1/2}\drm^n q) $  so that  the  inner  product  in the latter were induced
   by the scalar product (\ref{spr}).
 In the  generic \rin, map (\ref{map})     can be constructed  only  as the quasi-non-relativistic asymptotic(i.e. for
  $c^{-2}  \rightarrow 0 $).
   In \cite{Tag2}, the space $\Phi_N^- \{S_\Sg\}$ of the following
asymptotic in $c^{-2}$  solutions of eq.(\ref{eq}) is  taken as $\Phi^-$:
\begin{equation}
\varphi(x;\,N ) \,= \,\sqrt{\frac{\h}{2mc}}\ exp\left(-\im \frac{mc}{\h}\, S_\Sg(x)\right) \hat V(x;\, N)
\psi(x;\,N ),\quad
 N = 0, 1, \dots  .      \label{az}
\end{equation}
The  objects $\Sg$,\  $S_\Sg$,\  $\psi$, and $ \hat V(x) $  are:
 \begin{itemize}
  \item $\Sg$ \emph{is  a given   Cauchy  hypersurface} in $V_{1,n}$ as defined by eq. (\ref{sg});
 \item $S_\Sg $ is a solution of the Hamilton--Jacobi equation $\ptl_\alpha S_\Sg \,\ptl^\alpha S_\Sg \,=\,1,$
 with  the  initial conditions
$\left.S_\Sg (x)\right|_\Sg\, = \, 0 $;  any  hypersurface $ S_\Sg(x) =  const $ forms a  level surface of the
normal geodesic flow  through $\Sg $ which plays the  role  of  proper  frame  of  reference  for the  quantum
particle  under  consideration;
  \item $\psi(x;\,N )$ is   \emph{ a solution  of the \Sch equation}
  \begin{eqnarray}
\quad i\h c(\ptl^\al S \ptl_\al&+&\frac12 \Box S )\psi (x;\,N ) \,
= \,\left( \hat H^{({\rm ft})}_N (x)+ \Oc\right)\, \psi (x;\,N ), \label{t1} \\
\hat H^{({\rm ft})}_N (x) &\defst&
  \hat H^{({\rm ft})}_0 (x)\,+\,\sum_{n=1}^N \frac{\hat h_n (x)}{(2mc^2)^n},
\label{t2} \\
  \hat H^{({\rm ft})}_0(x)\, &\defst& \,-  \frac{\h^2}{2m} \left(\Delta_S (x) -
\tilde\xi R_{\mathrm(g)}(x) + \left(\frac{1}{2}(\ptl^\al S \,\ptl_\al \Box S) \ +\, \frac{1}{4}
 (\Box S)^2 \right)\right); \label{h0}\\
   S  &\equiv& S_\Sg \quad \mbox{(here and in sequel for  simplicity)}; \label{S2}
\end{eqnarray}
 the superscript $({\rm ft})$ denotes the  field-theoretical origin  of  the  object.
   Operators $\hat h_n (x)$ are determined by recurrent relations starting with
   $\hat h_0 \equiv  \hat H^{({\rm ft})}_0$ ; their concrete form is not  essential for  purposes of
   the present paper because,  finally, it  will be  concentrated  on  exactly  non-relativistic case of $N = 0$.
  Wave functions $\psi(x; \,N ) \in  L^2 (S_\Sg; {\mathbb C};  \drm\sigma_S)$ ( $\drm\sigma_S$ being   defined  as in
  eq.(\ref{n}) with $\Sg \sim S_\Sg $)   in the  following asymptotic sense:
\begin{equation}
\{\varphi_1,\,\varphi_2 \}_S = \left(\psi_1,\,\psi_2\right)_S \,\defst \,\int_S \drm\sigma_S \,\ov\psi_1\, \psi_2
\, + \Oc, \quad \vp_1,\vp_2 \in \Phi_N^-\{S\} ;  \label{psi}
\end{equation}
 \item $\hat V(x; \, N)$ \emph{is a differential operator on } $ L^2 (S_\Sg; {\mathbb C};  \drm\sigma)$  the  particular
 form of  which is not important  in  sequel except  that $\hat V(x; \, N) = \hat 1 + O(c^{-2})$
 \end{itemize}
  All "hatted" operators act along the hypersurface
$ S \ni x $ that is they  are differential operators containing only the  covariant derivatives  $D_\alpha $ along
$S$.

Eq.(\ref{psi}) provides   $\Phi_N^-\{S\}$  with the structure of $L^2 (S; \Bbb C; \drm v_ S)$ and  $\psi$  by the
standard Born probabilistic interpretation in each configurational space $S =const$, i.e. $|\psi(x)|^2$  is  the
 probability density to observe  the field configuration  which  may be called "a particle" at  the point
 $x \in S $.  At least, this field configuration satisfies an  intuitive  idea of the quantum particle as a localizable
object.

 Further, let  $\check {\cal O} $ signifies  any  of the QFT-operators of  observables in the  Fock representation
 determined by the space
  $\Phi_N^-\{S_\Sg\}$, which have been  introduced  above  in  Section 3. Then, owing to   relation (\ref{z}),
  the corresponding  asymptotically Hermitean  quasi-non-relativistic  QM-operator $\hat O $  is determined  up
  to an  asymptotic unitary  transformation   by  the following general relation:
 \begin{eqnarray}
<\varphi_1|\check {\cal O}|\varphi_2>&=&\left(\psi_1 \hat O_N\psi_2\right)_S\defst\int_S \drm\sigma_S
\,\ov\psi_1\, \hat O_N \psi_2 \,
+ \Oc,  \vp_1,\vp_2 \in \Phi^-_N\{S\}  \label{vv}\\
\hat O_N &\defst& \hat O_0 +\,\sum_{l=1}^N \frac{\hat o_l}{(2mc^2)^l};
 \label{O1}
\end{eqnarray}
again, $\hat o_n$  are  differential QM-operators along $S$ determined by recurrence relations starting with
$(\hat O)_0$.
 The simplest example of the  relation is
\begin{equation}
<\varphi_1|\,  \check {\cal N} (\hat \varphi;\,\Sg)\, |\varphi_2> = \frac{\left(\psi_1,\, \psi_2\right)_{S_\Sg}}
{\left(\psi_1,\, \psi_1\right)_{S_\Sg}^{1/2} \left(\psi_2,\, \psi_2\right)_{S_\Sg}^{1/2}} + \Oc,  \label{n3}
\end{equation}
and hence  the operator of the number of particles $ \hat {\cal N} (\hat \varphi;\,\Sg) $ is represented in the
space $\Psi_N\{S_\Sg\}\sim L^2 (S_\Sg; {\mathbb C};  \drm\sigma_S)$  by the unity operator as it should be in
quantum mechanics of a single stable particle .

In the  same way, one  could  determine the asymptotic  QM-operators of  particle position  $\hat q^a(x)$ and of
projection of  momentum on  a vector  field $K^\al (x)$ acting on $\Psi_N \{S_\Sg\}$   and \emph{along the
hypersurface} $S_\Sg (x)=const$.  The   formulae in  their  generality are somewhat lengthy and  I refer for  them
to \cite{Tag2}. Instead,  having in view as the main aim, comparison of the present asymptotic structure of the
field-theoretic origin with  QM in [I] obtained by quantization of the conservative natural mechanics, I  give
here   a summary of the  operators  for   the case  when $V_{1,n}$  \emph{is  a globally static space-time}.  In
this case,  coordinates $x$  can  be chosen  as $\{x^a\} \sim \{t, q^a\}$  so  that   the metric  of $V_{1,n}$
acquires the form  (\ref{ds}),  $ S = ct $ and \beq
    R_{(g)} (x) \equiv R_{(\om)} (x). \label{RR}
\nde Then,  the  asymptotic expansions of the QM-operators  of  observables   can be  represented
 as the formal closed expressions \cite{Tag2}:
\begin{eqnarray}
\hat H^{({\rm ft})}_\infty &=&  mc^2 \left(\left(\mathbf{\hat 1} + \frac{2\hat H^{({\rm
ft})}_0}{mc^2}\right)^{1/2} - \mathbf{\hat 1} \right); \qquad  \hat H^{({\rm ft})}_0 = -\frac{\h^2}{2m}(\Delta_S -
\tilde\xi \, R_{(\om)});
\label{hinf}\\
 \hat V_{\infty} &=&
\left(\mathbf{\hat 1} + \frac{2 \hat H^{({\rm ft})}_0}{mc^2}\right)^{-1/4}; \label{vinf} \\
({\hat p}_K)_\infty (x)  &=& -\frac{i\h}{2}
 \hat V_\infty^{-1}\cdot (K^\al D_\al)\cdot \hat V_\infty  +
\frac{i\h}{2} \hat V_\infty \cdot (K^\al D_\al)^{\dagger} \cdot\hat V_\infty^{-1},\quad  (K^\al\ptl_\al S = 0);
  \label{pinf}\\
c\ (\hat p_{\ptl S})_\infty (x) &=& mc^2 \left(\mathbf{\hat 1} + \frac{2\hat H^{({\rm ft})}_0}{mc^2}\right)^{1/2},
\qquad ({\rm the\ energy\ operator});\label{einf}\\
(\hat q_S^{(i)})_\infty (x) &=& q_S^{(i)} (x)\cdot\mathbf{\hat 1}  + \, \frac{1}{2} \left[ [\hat V_{\infty},\
q_S^{(i)} (x)],\ \hat V_\infty^{-1}\right].  \label{qinf}
\end{eqnarray}
These formulae are of interest for  separate investigation when $c^{-1}> 0$. For  example,  it is  seen that
operators of coordinates  $\hat q_S^{(i)} (x) $  do not commute except the  case of $S  \sim E_n $ and
$q_S^{(i)}(x) \equiv y^a $, the Cartesian  coordinates. However, I shall not  dwell on these interesting questions
here and pass  directly to the non-relativistic QM    resulting  from  this asymptotic structure in  the  limit
$c^{-1} = 0 $.

 \section{Non-relativistic Quantum Mechanics generated by \\ Quantum Field Theory}

It is  seen  that the  expressions (\ref{hinf} -- \ref{qinf}) are invariant  as \emph{w.r.t.}  the point
transformations
 $x^\al \longrightarrow \tilde x^\al (x)$    as well as \emph{w.r.t.} the  choice  of  classical
position-type observables $q_S^{(i)}(x)  \longrightarrow \tilde q_S^{(i)}(x) $ generated
  by the chosen initial $\Sg$ .

The  expressions for quantum observables
 for $c^{-1}=0 $ in terms of  arbitrary  coordinates $q^a $   on foliums $S$  of $V_{1,n}$  are
the  following differential operators acting on $\psi(t,\,q) \in L^2 (\rnf; {\mathbb C}; \omega^{1/2} \drm^n q)$:
\begin{itemize}
\item the  Hamilton  operator  for \Sch equation
\beq
 \hat H^{({\mathrm ft})}_0 (q) = -\frac{\h^2}{2m}(\Delta_\mathrm{(\om)} (q)
  - \tilde\xi \, R_\mathrm{(\om)}(q)\cdot \mathbf{\hat 1}
 \label{hinf0}
 \nde
  \item the operator of  projection of  momentum on the vector field $K^a(q)$ on $ V_n $
 \beq
 {\hat p}_K (q)  = -i\h \left (K^a \nabla^{(\om)}_a  + \frac12  (\nabla^{(\om)}_a K^a) \right) \cdot \mathbf{\hat 1}
 \equiv -i\h\, \frac1{\om^{1/4}}\frac{\ptl}{\ptl q^a}\cdot\left(\om^{1/4}K^a \right)  \cdot \mathbf{\hat 1}.   \label{pinf0}
 \nde
  where  $\hat O_1 \cdot \hat O_2$ denotes the  operator  product of  these operators,   $\nabla^{(\om)}_a$ is the
 covariant  derivative in $S \sim V_n$ (i.e., i.e. \emph{w.r.t.} the metric tensor $\om_{bc}$) and
 $\om \defst \det\|\om_{bc}\|$;
  \item
  the  position  operator
  \beq
  \hat q^{(i)}(q)  \defst q^{(i)}(q) \cdot \mathbf{\hat 1}. \label{qinf1}
  \nde
 \end{itemize}
 Recall that  $q^{(i)}$ are  scalar  functions of $x^\al$ and,  thus, of $q^a$, which are    subordinated  to
  conditions (\ref{q}). Thus,  operators $\hat q^{(i)}(q), {\hat p}_K (q),  \hat H^{({\mathrm ft})}_0 (q),  $
  are  independent
  on choice of $q^a$ but depend  on choice of  scalars  $q{(i)}$ and  the vector field $K^a$.  In particular,
   $n$  vectors
   \beq
  {K^{(i)}}^a (q) \defst  \om^{ab}\frac{\ptl q^{(i)}}{\ptl q^b}. \label{Ki}
 \nde
form  a basis in the  tangent spaces of  $V_n$  determined by $q^{(i)}$. Then, if the values  of  the latter
scalars  are taken as coordinates $q^a$, i.e.
 \beq
          q^a \equiv q^{(a)}(q). \label{(a)}
     \nde
 Then, $ K^{(i)}_a = \delta^{(i)}_a$  and the  brackets in  the superscript  $(i)$ may be omitted. Finally,
 we come  Pauli's expression (12) in \cite{TagI} :
 \beq
 {\hat p}_a  = -i\h\, \frac1{\om^{1/4}}\frac{\ptl}{\ptl q^a}\cdot\om^{1/4}.\label{pinf2}
  \nde
 Though it looks  as  non-invariant operator \emph{w.r.t.} transformations of $q^a$,  actually  it
  is  tightly related to choice  of canonically conjugate  $q^a$  the  values of  which  are  fixed
  by   the  scalar  functions $q^{(i)}(x)|_\Sg $ and  cannot  be  transformed.  Thus, there is  no  sense
 to ask, is it  an 1-form  or  not. Actually,  it is  \emph{a  form-invariant}: if  we take another set
 of scalars $\tilde q^{(i)}$
 that formalizes  measurement  of  position in the configurational space $\Sg$ by  a complete set of operators ${\hat{\tilde q}^a}$,
 then other set of  momentum operators $ {\hat {\tilde p}_a}$ should be  taken in  the
 form  of  eq.(\ref{pinf2}). Consequently,  returning to  canonical   quantization as in Section 3 of \cite{TagI}   with  these  changed
basic observables   gives  different QP related to the the canged scalars $q^{(i)} (x)$ formalising observation of
a particle position on a folium $S$.

 \section{Conclusion}
Summarizing   the main results  of  the  both  papers we come  to  the  following logical chain.
\begin{itemize}
\item[1.] If the \Sch variational quantization procedure \cite{Sch1} is  revised so  that the  canonically conjugate primary
 quantum observables $\hat q^a, \hat p_b $  were Hermitean operators (condition  of observability),  then
  QP appears  in  the Hamilton operator,  which   paradoxically depend on choice of coordinates $q^a$,
  (see \cite{TagI}, Section 3).
\item[2.] Then,   it  was  natural  to  review and investigate other popular quantization procedures in
application to  the natural  systems. Actually, QP was discovered by  DeWitt (1952) in a particular  version of
canonical quantization and it remarkably coincides with the  revised version of  \Sch quantization. Another
versions of canonical  quantization, as well as  quasi-classical,  geometrical and  Feynman (path  integration)
quantizations also generate different  QPs  with the common property that at the origin of  quasi-Euclidean
coordinates $y^a$
 all these quantization generate QP of  the  form
\beq
   V^\mathrm{(qm)}(y) = - \frac{\h^2}{2m} \cdot \tilde\xi R_{\mathrm(\om)}(y) + O(y). \label{Vqmx}
 \nde
 Moreover,   the mentioned  latter three quantizations  as well as  the (revised) \Sch variational  and
 canonical  DeWitt quantizations  give
\beq
        \xi = \frac 16, \label{16}
 \nde
that is the formula (\ref{Vqmy} )

Generalization  of  the  canonical  quantization   general (\cite{TagI}, Section 4) can  give any  value of $\xi$
and  some form of non-invariant QP persists to  appear.
\item[3.] If QM of a  natural system considered as QM of  a  particle in an external  static  gravitational ($n$-dimensional)
 field presented general-relativistically  as $ V_{1,n} \sim R \times V_n$,  then the term  (\ref{Vqmx})in  the Hamiltonian
  may  be considered  as a  violation of  PE in  Weinberg's formulation, see \cite{TagI}, Section 3,
  if \Sch equation may  be considered as  "a law  of nature" assumed  by Weinberg.
 \item[4.] In  view  of  this discouraging features  of QP in the non-relativistic QM of natural systems, an
 alternative approach  to  construction   of  QM of  a particle in  the  generic Riemannian  space-time  $V_{1,n}$ has
 been  considered. It  starts with quantum theory of linear scalar field non-minimally coupled to  the metric
 with the arbitrary constant $\tilde\xi$   of non-minimality.
 \item[5.] Despite that there  are  a continuum of the Fock representations of  the quantum  field, the condition of  accord
 with PE of the  structure  of  singularities  of the causal Green functions (propagators)
 fixes uniquely the value  of $\tilde\xi$  just by eq.(\ref{16})for any space dimension $n$.  This value coincides
 with  the constant of conformal  coupling  $\tilde\xi\mathrm{(conf)} \equiv (n-1)/(4n)$ is  just for $n=3$ and \emph{our
      real space-time $V_{1,3}$ is  exceptional}   in this  sense.
\item[6.] Relation between  $\tilde\xi$ and $\xi$ from (\cite{TagI}) is  ascertained by extraction of
the non-relativistic  QM  in  $V_n$  from QFT in  our  alternative  approach. It is  done by determination of the
unique Fock representation  the one-quasi-particle sector of which simulate the  structure  of QMs  generated in
(\cite{TagI}) by quantization  of  the generic  natural system. The result is
\beq
       \tilde\xi = \xi,   \label{xi}
        \nde
though $\hat H^{({\mathrm ft})}_0 (q)$  differs  from hamiltonians $\hat H (q) $ in (\cite{TagI}) obtained by
quantization of the  natural  mechanics  by that the  former does not contain the part of  QP depending  on choice
of coordinates $q$, that is  the  terms that are  hid in the  residual term $O(y)$ in eq.(\ref{16}).
\item[7.] That $\tilde\xi = 1/6$  required by  PE and  Eq.(\ref{xi}) together  mean that
\emph{QP is not  an  artefact or   a mistake and    inevitable in the frameworks of the traditional
(non-relativistic) quantization  formalisms and the canonical  quantization of  general-relativistic  non-minimal
scalar field}.
\end{itemize}

Meanwhile,  there  is  a difference between QMs in  $V_{1,n}$ in  that quantization  of the natural systems
generates   a more complicate  QP which does not vanish even in  the  Euclidean  space-time  $E_{1,n}$  if
curvilinear coordinates  are  taken  as the  position observables $q^a$. I have  attempted in \cite{TagI} to
interpret this phenomenon as intervention  of  information  on the (speculative) classical position detecting
device. into  the quantum Hamiltonian. The relativistic  theory cannot  include information on  such a device in
principle and takes into account only  the  local QP in (\ref{hinf0}). The difference between  the two  approaches
is  not a discrepancy, in my opinion,  but different particular  manifestations of a more  deep quantum physics
still unknown for us completely but apparently related to  the  problem of  measurement. Recall also that some
essential considerations related  to  the problem are given  in the  last section of \cite{TagI}.
 \section{Acknowledgement}
 The  author is  thankful to  Professors P.~Fiziev, V.~V.~Nesterenko and  S.~M.~Eliseev  for useful
 discussions �nd consultations.

\end{document}